\documentclass[conference]{IEEEtran}
\IEEEoverridecommandlockouts

\usepackage[utf8]{inputenc}
\usepackage[T1]{fontenc}
\usepackage[american]{babel}
\usepackage[center]{caption}
\usepackage[hidelinks, urlcolor=black, colorlinks=true, linkcolor=black, citecolor=black  ]{hyperref}
\usepackage{bera}
\usepackage{graphicx}
\usepackage{float}
\usepackage{booktabs}
\usepackage{siunitx}
\usepackage{amsmath,amssymb,exscale}
\usepackage{blindtext, graphicx}
\usepackage{verbatim}
\usepackage{algpseudocode}
\usepackage{fancyvrb}
\usepackage{amsmath}
\usepackage{mathtools}
\usepackage{lipsum}
\usepackage{acronym}\acrodef{3GPP}[3GPP]{3rd generation partnership project}
\acrodef{AI}[AI]{artificial intelligence}
\acrodef{WI}[WI]{Wireless InSite\textsuperscript{\textregistered}}
\acrodef{IP}[IP]{Internet Protocol}
\acrodef{B5G}[B5G]{beyond 5G}
\acrodef{KPI}[KPI]{key performance indicator}
\acrodef{5G}[5G]{5th generation}
\acrodef{SLA}[SLA]{service-level agreement}
\acrodef{MAPE}[MAPE]{Mean Absolute Percentage Error}
\acrodef{5G/6G}[5G/6G]{5th and 6th generation}
\acrodef{QoS}[QoS]{quality of service}
\acrodef{LoRaWAN}[LoRaWAN]{long range wide area network}
\acrodef{LPWAN}[LPWAN]{low power wide area network}
\acrodef{IoT}[IoT]{internet of things}
\acrodef{RT}[RT]{ray tracing}
\acrodef{DT}[DT]{digital twin}
\acrodef{NDT}[NDT]{network digital twin}
\acrodef{ITU}[ITU]{International Telecommunication Union}
\acrodef{MLP}[MLP]{multilayer perceptron}
\acrodef{UDP}[UDP]{User Datagram Protocol}
\acrodef{TCP}[TCP]{Transmission Control Protocol}
\acrodef{PDR}[PDR]{packet delivery ratio}
\acrodef{SF}[SF]{spreading factor}
\acrodef{ED}[ED]{end-device}
\acrodef{GW}[GW]{gateway}
\acrodef{UAV}[UAV]{unmanned aerial vehicle}
\acrodef{CDF}[CDF]{cumulative distribution function}
\acrodef{MILP}[MILP]{mixed integer linear programming}
\acrodef{BILP}[BILP]{binary integer linear programming}
\acrodef{MBNLP}[MBNLP]{mixed binary non-linear programming}
\acrodef{MNLP}[MNLP]{mixed non-linear programming}
\acrodef{MWF}[MWF]{multi-wall and multi-floor}
\acrodef{ILP}[ILP]{integer linear programming problem}
\acrodef{GA}[GA]{genetic algorithm}
\acrodef{RMSE}[RMSE]{root mean squared error}
\acrodef{EPSO}[EPSO]{evolutionary particle swarm optimization}

\usepackage{scalerel}
\usepackage{tikz}
\usetikzlibrary{svg.path}

\definecolor{orcidlogocol}{HTML}{A6CE39}
\tikzset{
  orcidlogo/.pic={
    \fill[orcidlogocol] svg{M256,128c0,70.7-57.3,128-128,128C57.3,256,0,198.7,0,128C0,57.3,57.3,0,128,0C198.7,0,256,57.3,256,128z};
    \fill[white] svg{M86.3,186.2H70.9V79.1h15.4v48.4V186.2z}
                 svg{M108.9,79.1h41.6c39.6,0,57,28.3,57,53.6c0,27.5-21.5,53.6-56.8,53.6h-41.8V79.1z M124.3,172.4h24.5c34.9,0,42.9-26.5,42.9-39.7c0-21.5-13.7-39.7-43.7-39.7h-23.7V172.4z}
                 svg{M88.7,56.8c0,5.5-4.5,10.1-10.1,10.1c-5.6,0-10.1-4.6-10.1-10.1c0-5.6,4.5-10.1,10.1-10.1C84.2,46.7,88.7,51.3,88.7,56.8z};
  }
}

\newcommand\orcid[1]{\href{https://orcid.org/#1}{\mbox{\scalerel*{
\begin{tikzpicture}[yscale=-1,transform shape]
\pic{orcidlogo};
\end{tikzpicture}
}{|}}}}

\def\BibTeX{{\rm B\kern-.05em{\sc i\kern-.025em b}\kern-.08em
    T\kern-.1667em\lower.7ex\hbox{E}\kern-.125emX}}

\makeatletter 
\let\old@ps@headings\ps@headings 
\let\old@ps@IEEEtitlepagestyle\ps@IEEEtitlepagestyle 
\def\confheader#1{%
\def\ps@headings{%
\old@ps@headings%
\def\@oddhead{\strut\hfill#1\hfill\strut}%
\def\@evenhead{\strut\hfill#1\hfill\strut}%
}%
\def\ps@IEEEtitlepagestyle{%
\old@ps@IEEEtitlepagestyle%
\def\@oddhead{\strut\hfill#1\hfill\strut}%
\def\@evenhead{\strut\hfill#1\hfill\strut}%
}%
\ps@headings%
} 
\makeatother 

\begin{document}

\title{Ray Tracing-Based LoRaWAN Gateway Placement for Reliable Connectivity in Amazonian Regions}

\author{\IEEEauthorblockN{Cláudio Modesto \orcid{0009-0005-5184-8030
}\IEEEauthorrefmark{1}, Lucas Mozart \orcid{0009-0000-2092-4399}\IEEEauthorrefmark{1}, 
Cleverson Nahum \orcid{0000-0001-9644-5394}\IEEEauthorrefmark{1},\\
Bruno Castro \orcid{0000-0003-4601-3205}\IEEEauthorrefmark{1}, 
Aldebaro Klautau \orcid{0000-0001-7773-2080
}\IEEEauthorrefmark{1}}
\IEEEauthorblockA{LASSE, Federal University of Pará\IEEEauthorrefmark{1}\\
Email: claudio.barata@itec.ufpa.br\IEEEauthorrefmark{1}}

}

\IEEEoverridecommandlockouts 
\IEEEpubid{\makebox[\columnwidth]{Copyright Notice } 
\hspace{\columnsep}\makebox[\columnwidth]{ }} 

\maketitle

\begin{abstract}
Network planning is an important task in wireless communications, as it helps network operators avoid unnecessary costs. In the context of the internet of things, using long-range wide-area network technologies in the Amazon rainforest, a key challenge is ensuring reliable communication between end-devices and gateways (GWs). In this sense, this reliability is strongly affected by channel conditions. Thus, during the planning phase, choosing the appropriate channel model is an important decision for accurate simulations.
Given this motivation, in this work, we propose an optimization model to evaluate the impact of different types of channels on coverage and packet delivery ratio in a forest scenario. We used channels from ray tracing, empirical, and stochastic approaches to assess how decisions made during the network planning phase, in terms of the channel used, affect GW placement and, specifically, the percentage of end-devices covered and the reliability of the communication system. Our results show that GW placement based on site-independent channels can overestimate the number of GWs required to meet the network requirements, whereas using site-specific channels allows us to satisfy the same requirements with fewer GWs.
\end{abstract}


\section{Introduction}
Network planning represents an important aspect of wireless network communication in the phases before the actual deployment of the network~\cite{Biswajit2024}. For certain regions, such as the Amazon rainforest sites, this type of task is essential to ensure reliable connectivity for several applications, such as \ac{IoT} devices used in fauna and flora tracking applications related to wildlife conservation~\cite{ross2022, Wild2023}. Under this motivation, a specific task related to network planning in \ac{LoRaWAN} communications is the positioning of \acp{GW}. In this sense, this kind of problem is extremely dependent on the type of channel used to model the propagation effects inherent to this type of communication, such that it is possible to use site-independent approaches represented by stochastic~\cite{3gppTR38901} and empirical channels~\cite{Correia2023, Pires2024}, or site-specific channels~\cite{modesto2026} obtained from \ac{RT} simulators, such as Sionna or \ac{WI}~\cite{wirelessInsite}.

Considering these several ways to model the physical propagation effects of this type of communication, the problem of \ac{LoRaWAN} \acp{GW} positioning has been widely investigated in the literature~\cite{Correia2023, mahtre2024, said2025, Matni2020}, with each work adopting specific objectives regarding \ac{GW} positioning. Nevertheless, most converge on two common characteristics: a focus on macro-urban scenarios and the use of site-independent channel models mentioned earlier. Therefore, challenges related to forest environments and the specific effects of the site in the propagation model are simplified. To address this gap, we extend our previous work~\cite{modesto2026} and investigate the impact of different channel models in Amazon forest-like \ac{LoRaWAN} scenarios.
 
Therefore, our contributions to this paper are:

\begin{itemize}
    \item An optimization model for \ac{GW} placement in forest scenarios using \ac{RT}-based channels.
    \item An analysis of the impact of channel model choices on \ac{PDR} and coverage metrics in the context of network planning in forest scenarios.
    \item Open artifacts for the community containing a highly detailed forest 3D scenario and a dataset with large scale parameters obtained with a \ac{RT} simulator.\footnote{Source code is available at: \url{https://github.com/lasseufpa/forest-rt}}
\end{itemize}

The remainder of this paper is organized as follows: Section \ref{sec: gateway_placement} describes the \ac{GW} placement problem considered in this work, with details on the objective function to be optimized, the decision variables, and constraints. Section~\ref{sec: experiments} details the proposed experiments to evaluate the impact of different channel models on \ac{GW} position in forest scenarios. Using this proposed experiment, Section~\ref{sec:results} discusses the main results obtained. Finally, Section \ref{sec:conclusions} concludes this work and outlines its main results.

\section{Gateway placement optimization}
\label{sec: gateway_placement}

We initially assume that each \ac{GW} can be placed in the position index $p$, which belongs to the set of candidate locations $\mathcal{P} = \{1, 2, \dots, p, \dots, P\}$, where each element corresponds to a Cartesian coordinate on the grid, and $P$ denotes the total number of candidate positions. We also assume that an \ac{ED} $d$ comes from a set $\mathcal{D} = \{1, 2, \dots, d, \dots, D\}$ consisting of a total of $D$ devices. Furthermore, each \ac{ED} can use the $s$-th \ac{SF} from a set of \ac{SF} configurations $\mathcal{S}$. With these sets, we formally formulate the problem addressed in this work as a \ac{BILP} with three binary decision variables $x^{p}$, $a^p_d$, and $y_s^p$. $x^{p}$ indicates whether a \ac{GW} is installed at position $p$. $a^p_d$ indicates whether the \ac{GW} in position $p$ is serving the $d$-th \ac{ED}. Finally, $y_s^p$ indicates if  the $p$-th \ac{GW} is operating with $s$-th \ac{SF}. 

Each possible \ac{GW} position is associated with an aggregated \ac{PDR} without collisions $\text{PDR}_{\text{no\_collision}}$, such that
\begin{equation}
\text{PDR}_{\text{no\_collision}}
=
[f(1), f(2), \dots, f(p), \dots, f(P)],
\end{equation}
where $f : \mathcal{K} \rightarrow [0,1]$ is a function that maps a deployed \ac{GW} position to its corresponding aggregate $\text{PDR}_{\text{no\_collision}}$. In this case, we used a \ac{PDR} without collisions because computing all the \ac{GW} interactions and their resulting collisions in all possible \ac{GW} positions was too computationally demanding. Therefore, considering this system model, our goal is to minimize the number of deployed \ac{GW}, leading to the following objective function:
\begin{equation}
\min_{x^{p}} \sum_{p \in \mathcal{P}} x^{p}.
\label{eq: objective_function}
\end{equation}

Let $\rho$ denote the received power threshold for each \ac{ED}, and let $\omega_d$ represent the received power achieved by the $d$-th \ac{ED} after the \ac{GW} is positioned. Then, the first constraint of the objective function in Eq.~(\ref{eq: objective_function}) is intended to ensure that a certain percentage of the total \ac{ED} is covered and consists of multiple received-power coverage constraints, one for each \ac{ED}. This can be expressed as follows:
\begin{equation}
    \sum_{d \in \mathcal{D} }\left\lceil\sum_{p\in \mathcal{P}} \dfrac{\beta_d^p x^{p}}{|\mathcal{P}|}\right\rceil \geq \alpha |\mathcal{D}|,
    \label{eq: power_constraint}
\end{equation} 
\noindent which $\alpha$ indicates the percentage of \ac{ED} that should be covered,  $\beta_d^p = 0$ indicates that the $d$-th \ac{ED} is not covered by the \ac{GW} in the $p$-th position ($\omega_d < \rho$), and $\beta_d^p = 1$ indicates that the \ac{ED} is covered by the \ac{GW} at the $p$-th position ($\omega_d \ge \rho$). 

 
The second set of constraints is related to the \ac{PDR} achieved from the positioned \acp{GW}.  Therefore, knowing that $\gamma$ is the minimum \ac{PDR} threshold, we define the \ac{PDR} constraint as follows:
\begin{equation}
     \sum_{s \in \mathcal{S}}\zeta^p_s y_s^p =  x^p, \quad \forall p\in \mathcal{P},
    \label{eq: pdr_constraint}
\end{equation}
which $\zeta_s^p$ is a binary parameter that indicates whether the \ac{GW} positioned in the $p$-th location is providing a sufficient \ac{PDR} $\Omega_p$ to all \acp{ED} that satisfy the minimum requirements, with $\zeta_s^p=0$ not satisfying ($\Omega_p < \gamma $) and $\zeta_s^p=1$ satisfying the requirement ($\Omega_p \ge \gamma $). Finally, the last two sets of constraints concern the uniqueness of the \ac{GW}–\ac{ED} link, ensuring that each \ac{ED} is covered by at most one \ac{GW} and only by a selected \ac{GW} that provides coverage. Formally, we can define this as:
\begin{equation}
    \sum_{p \in \mathcal{P}}a_d^p \le 1, \quad \forall d \in \mathcal{D},
\end{equation}
\noindent and
\begin{equation}
    a_d^p \le  \beta_d^px_d^p, \quad \forall d \in \mathcal{D}, \forall p \in \mathcal{P}.
\end{equation}
\section{Experiments}
\label{sec: experiments}
\begin{figure}[!h]
    \centering
    \includegraphics[scale=0.22]{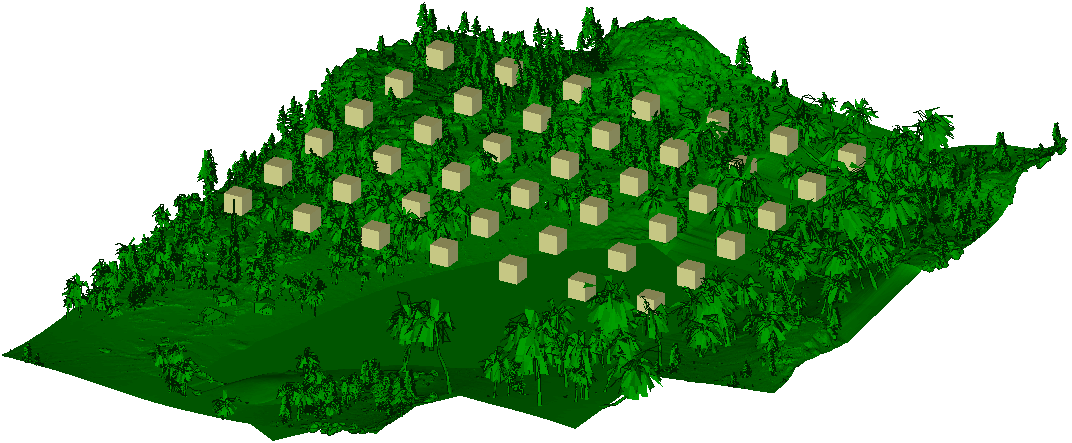}
    \caption{3D forest scenario used in the proposed experiments. The cubes in the scenario represent the possible \ac{GW} positions.}
    \label{fig:3d_scenario_used}
\end{figure}

Intended to evaluate the impacts of the channel model choices on the \ac{GW} placement optimization in a forest scenario, we propose several experiments in a rural scenario, as defined in Fig.~\ref{fig:3d_scenario_used}. This 3D scenario can be characterized by a highly detailed geometry, with a total of 870\,008 faces. Moreover, this scenario is predominantly composed of vegetation, with the following electromagnetic parameters: the relative permittivity equal to 1.1 and conductivity equal to 0.5~\cite{tewari1986}.

\begin{figure}[!h]
    \centering
\includegraphics[scale=0.48]{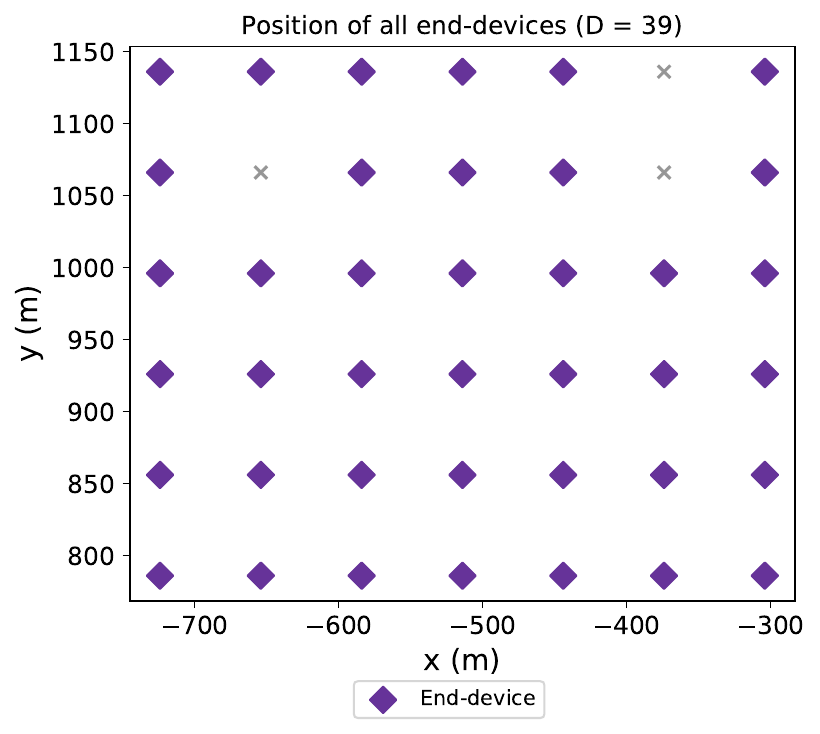}
    \caption{Upper view perspective of spatial organization of \ac{ED} in the 3D scenario.}
    \label{fig:ed_spatial_organization}
\end{figure}

For the \ac{GW} placement optimization problem, we considered the $6 \times 7$ grid of possible \ac{GW} positions. For each possible position, a \ac{GW} is deployed in a height of 17 \si{m}. The positioned \acp{GW} should serve the set of $\mathcal{D}$, considering the following spatial organization depicted in Fig.~\ref{fig:ed_spatial_organization}. Each \ac{ED} has a height of 2 \si{m}.

For all experiments, we consider two groups of channel models. The first group consists of site-independent channel models composed of Okumura-Hata, 3GPP-RMa~\cite{3gppTR38901}, log-distance, and COST-231. The second group consists of \ac{RT}-based channel models obtained from \ac{WI} using the X3D and Full-3D \ac{RT} algorithms. For the site-independent channels, we configured these models considering the log-distance model with a path-loss exponent of $3.76$ and a reference distance of $32$ \si{m}. For the frequency-dependent models (excluding the log-distance model), we used a carrier frequency of 1 \si{GHz}. Finally, for the models that incorporate the heights of the \ac{GW} and \ac{ED}, we used the previously specified values of 17~\si{m} for the \ac{GW} and 2~\si{m} for the \ac{ED}. For the site-specific channel model obtained from \ac{RT} simulator, we considered the parameters defined in Table~\ref{tab:rt_parameters}.

\begin{table}[htp]
    \centering
    \caption{\ac{RT} simulation parameters used for the experiments with site-specific channels using \ac{WI}}
    \scalebox{1}{\begin{tabular}{lc}
    \toprule
      Simulation parameter & Value\\
    \midrule
       Carrier frequency (\si{GHz})  & $1$ \\
       Radiation pattern & Isotropic \\
       Tx polarization & Vertical \\
       Maximum number of interactions & 6\\
       Tx waveform & Sinusoid\\
       Interactions & Diffraction and reflection \\
\bottomrule
    \end{tabular}}
\label{tab:rt_parameters}
\end{table}

Using these configurations, the first experiment considers a spatial analysis of how different the \ac{GW} positioning is when we consider different channel models. For the thresholds, we assumed the minimum received power $\rho=-95$~\si{dBm}, the minimum \ac{PDR} $\gamma=0.7$, and the minimum percentage of \ac{ED} covered $\alpha=0.8$. The second experiment is a sensitivity analysis related to the minimum received power ($\rho$) threshold in the optimization model. Finally, we conducted these simulations considering the \ac{RT}-\ac{LoRaWAN} framework from~\cite{modesto2026} in a configured environment, as defined by Table~\ref{tab:configuration_setup}. In this environment, we primarily used tools such as \ac{WI} and ns-3~\cite{henderson2008network}. For ns-3 simulations, we utilized a server equipped with an Intel\textsuperscript{\textregistered} Xeon Silver 4514Y, a NVIDIA\textsuperscript{\textregistered} A2 GPU, and 512 \si{GB} of RAM. For \ac{RT} simulations with \ac{WI}, we used a server equipped with an Intel\textsuperscript{\textregistered} Core\textsuperscript{\texttrademark} i7-10700F CPU @ 2.90 GHz, a NVIDIA\textsuperscript{\textregistered} GeForce 3060 GPU, and 64 \si{GB} of RAM.

\begin{table}[!h]
    \centering
    \caption{Setup configuration used to generate data and solve the proposes \ac{GW} placement optimization}
    \scalebox{1}{\begin{tabular}{lc}
    \toprule
    Tool  & Version \\
    \midrule

    Operating system & Ubuntu 20.04.6 and Windows 10 \\
    Pyomo & 6.9.5   \\
    ns-3 & 3.45 \\
    GLPK & 4.65\\
    Wireless InSite & 3.4.5 \\
    \bottomrule
    \end{tabular}}
    \label{tab:configuration_setup}
\end{table}

\section{Experimental Results}
\label{sec:results}
\begin{figure}[!h]
    \centering
\includegraphics[scale=0.5]{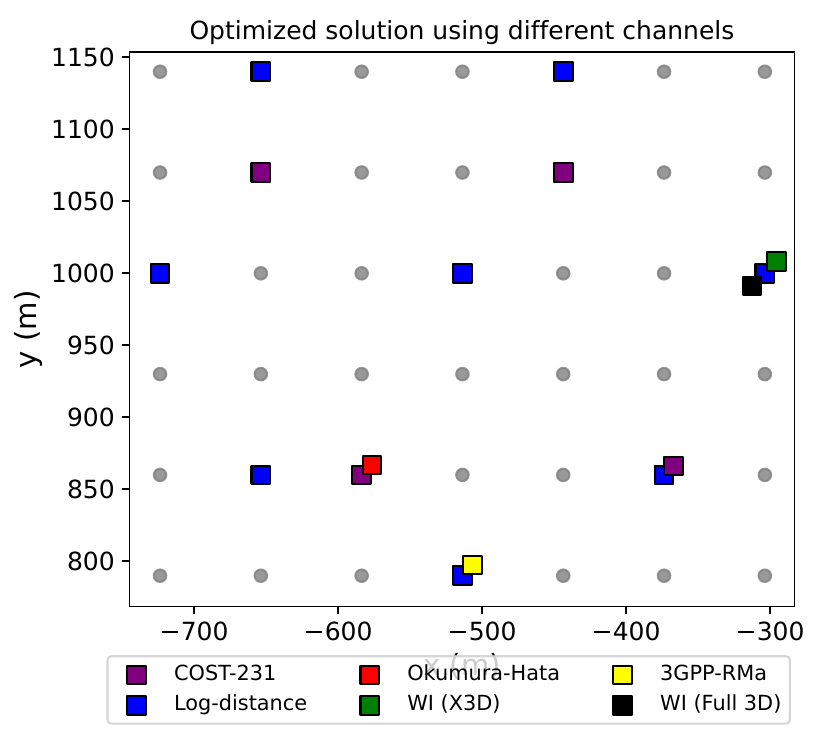}
    \caption{Optimized solution for \ac{GW} placement in the forest scenario, considering site-specific and site-independent channels.}
    \label{fig:result_positions}
\end{figure}

For the first experiment, the spatial analysis of the optimization result, depicted in Fig.~\ref{fig:result_positions}, shows high sensitivity in the number of suggested \acp{GW} across site-specific and site-independent channel models.

\begin{figure}[!h]
    \centering
    \includegraphics[scale=0.49]{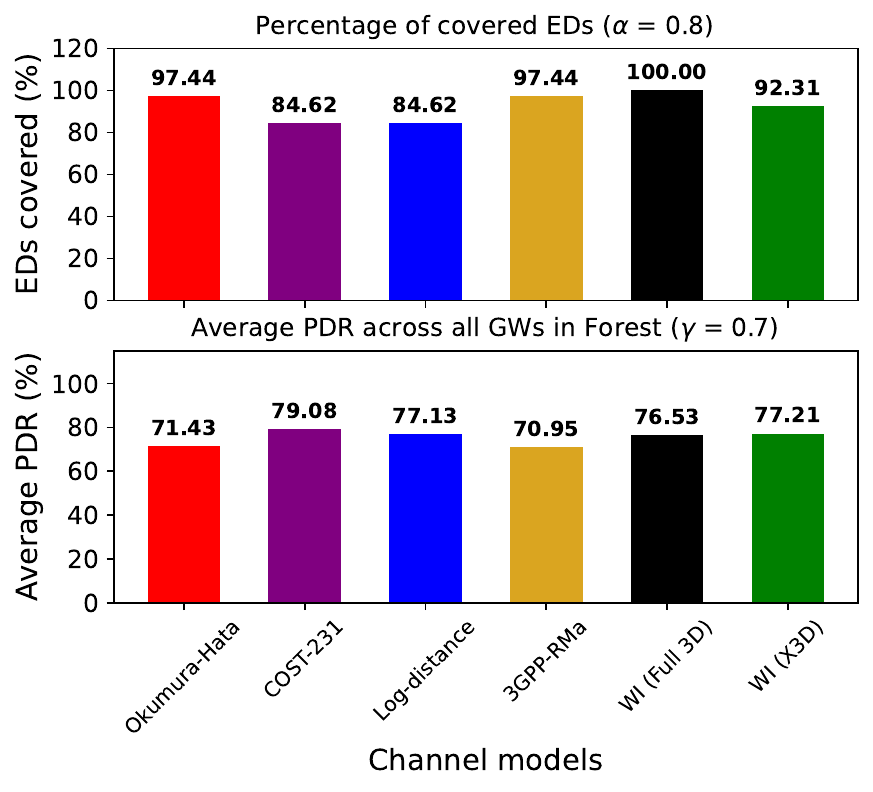}
    \caption{Percentage of covered \ac{ED} (upper) and average \ac{PDR} (bottom), considering solutions using different channel models.}
    \label{fig:eds_covered}
\end{figure}

In this spatial analysis, the number of deployed \acp{GW} can vary significantly depending on the channel model. For example, with the COST-231 channel, the number of \acp{GW} is 4, whereas with the log-distance model, it increases to 8. Using Okumura-Hata or 3GPP-RMa, the number of suggested \acp{GW} is only 1, but in different positions. For the site-specific channels, starting \ac{WI} X3D and Full 3D, the number of \acp{GW} remains the same in equal positions. In this analysis, we observe that the position and number of \acp{GW} obtained using the Log-distance and 3GPP-RMa channel models are close to each other (likewise the solutions using WI X3D and Full 3D). However, although similar optimized solutions are applied with these channel models, their impact is not uniform. When considering the percentage of covered \ac{GW} and the \ac{PDR}, we can see the difference between both solutions, as depicted in Fig.~\ref{fig:eds_covered}.

From Fig.~\ref{fig:eds_covered}, we observe that, according to the log-distance model, the percentage of covered \acp{ED} remains below 85\%, whereas the 3GPP-RMa and Okumura-Hata models predict coverage of nearly 98\%. For the site-specific channels, even when the \acp{GW} are placed at the same locations, the resulting coverage percentages differ, with the solution using \ac{WI} Full 3D achieving full coverage. Furthermore, when we incorporate the \ac{PDR} into the analysis, the impact of the channel characteristics becomes more evident, allowing us to quantify the cost of achieving a given \ac{PDR} in terms of the number of \ac{GW}. In this context, models such as Log-distance and COST-231 require a significantly higher number of \ac{GW} to achieve similar \ac{PDR} as that obtained using \ac{RT} channels, with just one \ac{GW}. This clear difference between site-specific and site-independent channel models can be particularly problematic for network planning in forest environments and highlights the importance of considering channels obtained from specific sites. In this sense, this difference in each result can bias the decisions of network operators regarding deployment, hiding possible connectivity issues and increasing unexpected maintenance costs after the deployment process.

\begin{figure}[!h]
    \centering
    \includegraphics[scale=0.45]{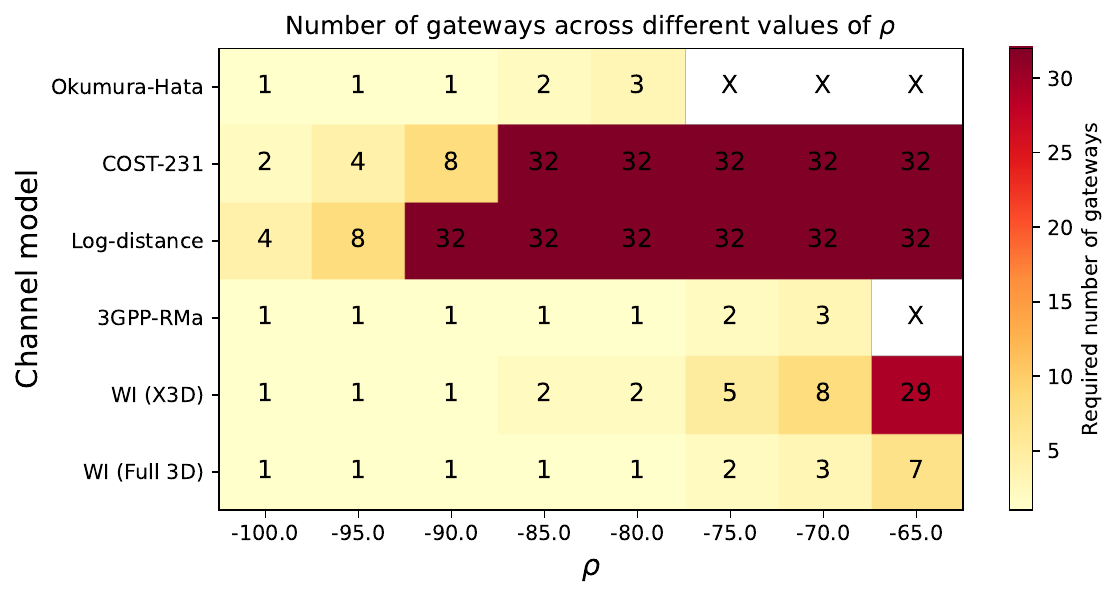}
    \caption{Sensitivity analysis considering values of $\rho$ from $-100$ to $-65$ \si{dBm}, $\alpha =0.8$, and $\gamma=0.7$.}
    \label{fig:sensit_rec_power}
\end{figure}

For the second set of evaluations, we have a sensitivity analysis considering different values of $\rho$. In this sense, Fig.~\ref{fig:sensit_rec_power} depicts the number of \acp{GW} for each value of $\rho$ with fixed thresholds for the minimum number of covered devices and \ac{PDR}. From this result, we can detect that for some channel models, the optimization process presents no feasible solution. This is the case for the Okumura-Hata and 3GPP-RMa models.

Another important aspect is the striking difference between site-specific and site-independent models in  the number of required \acp{GW} for some values of $\rho$. A remarkable result is the ratio of \acp{GW} to \acp{ED} required to achieve a minimum received power of \(-85\)~\si{dBm}. While the log-distance and COST-231 models each require 32 \acp{GW}, which is essentially an unrealistic deployment of almost one \ac{GW} per \ac{ED}, the \ac{WI} X3D and \ac{WI} Full 3D solutions require only 2 and 1 \acp{GW}, respectively.

\section{Conclusions}
\label{sec:conclusions}
In this paper, we proposed an optimization model for \ac{GW} placement in amazonian environments that incorporates different types of channel models, including \ac{RT}-based, empirical, and stochastic approaches. For the empirical and stochastic channels, we constructed a network environment with a pre-defined position grid to allocate all \acp{GW} and \acp{ED}. For the site-specific channel, we used the same position grid in a 3D scenario that included trees and vegetation, explicitly accounting for their electromagnetic properties. Using these environments, we conducted several experiments that reveal a strong sensitivity of the optimized solutions to the chosen channel model. In our spatial analysis, we observed that even when \acp{GW} are placed at the same locations, employing different channel models can result in significantly different percentages of covered \acp{ED} and \ac{PDR}. Furthermore, in the sensitivity analysis of the received power threshold, we detected that some site-independent channel models can present infeasibility issues that are not present in the site-specific channels obtained from \ac{WI}. Finally, from this sensitivity analysis, we also show that the optimization process with some site-independent channels can generate unrealistic solutions, suggesting, for example, almost one \ac{GW} per \ac{ED} to satisfy the coverage constraints.

\section*{Acknowledgment}
This study was financed in part by the Coordenação de Aperfeiçoamento de Pessoal de Nível Superior - Brasil
(CAPES), in part by the Conselho Nacional de Desenvolvimento Científico e Tecnológico (CNPq); the Brasil 6G project (01245.020548/2021-07), supported by RNP and MCTI; in part by the Innovation Center; in part by the Project Smart 5G Core And
MUltiRAn Integration SAMURAI (MCTIC/CGI.br/FAPESP under Grant 2020/05127-2); and in part by the National Institute of Science and Technology (INCT) STREAM, funded by the CNPq, process no. 409179/2024-8.

\bibliographystyle{IEEEtran}
\bibliography{references}

\end{document}